\documentstyle[twocolumn,aps]{revtex}
\begin{document}

\title{Third Newton Law in the Mean  and Perturbations of Noise}

\author{Piotr Garbaczewski\\
Institute of Theoretical Physics, University of Wroc{\l}aw, \\
pl. M. Borna 9,  PL-50 204 Wroc{\l}aw, Poland\\
 and \\
 Institute of Physics, Pedagogical University,\\
  pl. S{\l}owia\'{n}ski 6, PL-65 069 Zielona G\'{o}ra,
  Poland\thanks{Email:  pgar@omega.im.wsp.zgora.pl}}
\maketitle
\centerline{PACS numbers:  02.50-r, 05.20+j, 03.65-w}
\begin{abstract}
We analyze an  isothermal Brownian motion
of particle ensembles in the Smoluchowski approximation.
 Contrary to standard procedures,
the environmental recoil effects associated with locally
induced heat flows are not completely disregarded.
The main technical tool in the study of such  weakly-out-of
equilibrium  systems is a consequent  exploitation of the
Hamilton-Jacobi equation.
The third Newton law in the mean is utilised to generate
diffusion-type processes which are either anomalous
(enhanced), or generically non-dispersive.
\end{abstract}
\vskip2cm

If we consider  a   fluid in thermal equilibrium as the noise
 carrier, a kinetic theory viewpoint  amounts to visualizing the
 constituent molecules that collide
not only with each other but  also  with the tagged (colloidal)
particle, so \it enforcing \rm  and maintaining
its  observed erratic  motion.
The Smoluchowski \it approximation \rm takes  us   away
 from those kinetic theory intuitions by  projecting  the
phase-space theory of random motions into its  configuration
space image  which is a spatial Markovian diffusion
process:
$${d\vec{X}(t)= {\vec{F}\over {m\beta }}dt +
\sqrt{2D}d\vec{W}(t)\enspace .}\eqno (1)$$

In the above $m$ stands for the mass of a diffusing particle,
$\beta $ is a friction parameter, D is a diffusion constant
 and $\vec{W}(t)$ is a normalised Wiener process.
The Smoluchowski forward drift can be traced back to
a presumed selective action of the external force $\vec{F}=
-\vec{\nabla }V$ on the Brownian particle
that  has a negligible  effect on the thermal bath but
in view of frictional  resistance imparts to a particle
the \it  mean \rm velocity $\vec{F}/m\beta $ on
the $\beta ^{-1}$ time scale, \cite{smol,nel}.

Smoluchowski diffusions are conventionally
regarded as \it isothermal  \rm processes.
If we however admit that
tiny heat flows should  accompany  the particle
transport due to the Brownian motion, a suitable description
of thermal inhomogeneities and their   effects on the
dispersion of Brownian particles must be invented.

To exemplify this point, let us consider the
phenomenological assumption of Ref. \cite{streater}:
"a gas of  Brownian  particles falling
in gravity should leave a trail of warm fluid in its wake,
since its potential energy is being converted into heat".
Obviously,  if the  particles would move against
gravitational force,
then the temperature of the medium  should locally drop down.
Those features, if we are to keep track of the local heating and
cooling (as opposed to the isothermal Einstein or Smoluchowski
diffusive dynamics) were interpreted in Ref. \cite{streater}
as a source of the space-time
dependence of temperature.
In that case the Fokker-Planck equation must
be supplemented by  an evolution equation for the temperature
field.

The phase-space scenario usually refers  to
  minute acceleration/deceleration events  which
modify (say,  at a rate of $10^{21}$   per second)
velocities of  realistic    particles. Clearly,  the microscopic
energy-momentum  conservation laws need to be respected in
each separate collision event.  In contrast to derivations based
on the Boltzmann colllision scenario, this feature is completely
\it  alien \rm to the Brownian motion theory,
cf. \cite{vigier,blanch}.
Therefore,
there seems tempting to require that
each minute acceleration of a Brownian particle   is accompanied
by a minute \it cooling \rm of the medium in its immediate
neighbourhood, while any deceleration event should
induce a local \it heating  \rm phenomenon,  see
e.g. \cite{perrin}.     Since (cf. Eq. (1)), we
always disregard the fine details of about $10^{13}$ collision
impacts on the Brownian particle on a typical relaxation
time scale of $10^{-8}s$, there is  definitely enough  room to
allow for local  statistical measures of heating and cooling.

     In  a weakly out-of-equilibrium system
 with small heat flows, we may  expect that the
 standard equilibrium temperature notion needs a replacement
 by an \it effective \rm
 temperature notion (and an effective thermal equilibrium),
  which depends     on the chosen fast-versus-slow-process
  time scales and the ensemble averaging.
In the above sense only  an effective isothermal regime
may be maintained.

It is well known that
a spatial diffusion (Smoluchowski) approximation of the
phase-space process, allows
to reduce the number of \it independent \rm
local conservation laws
(cf. \cite{vigier,blanch,zambrini,geilikman}) to two only.
Therefore the Fokker-Planck  equation can
always be supplemented  by another (independent) partial
differential equation to form a \it closed \rm system.
Non-isothermal flow description needs
to accomodate variations of temperature of the bath,
(cf. \cite{streater}), while we investigate the limits of
validity of the isothermal  scenario.
That amounts to \it inequivalent \rm
choices of the supplementary equation.

 If we assign
 a probability density $\rho _0(\vec{x})$ with which the
 initial data
  $\vec{x}_0=\vec{X}(0)$ for Eq. (1) are
distributed, then the emergent Fick law  would  reveal a
statistical tendency of particles to flow away from
higher probability  residence areas.     This
feature is encoded in the corresponding Fokker-Planck equation:
$${\partial _t \rho  = - \vec{\nabla }\cdot (\vec{v}\rho )=
- \vec{\nabla }\cdot [({\vec{F}\over {m\beta }}  - D
{{\vec{\nabla }\rho }\over {\rho }}) \rho ]}\eqno (2)$$
where a diffusion current velocity  is
$\vec{v}(\vec{x},t) = \vec {b}(\vec{x},t) - D{{\vec{\nabla }\rho
(\vec{x},t)}\over {\rho (\vec{x},t)}}$
while  the forward drift reads $\vec{b}(\vec{x},t) =
{\vec{F}\over {m\beta }}$.
Clearly, the local diffusion current (a local flow that might
be experimentally observed for  a cloud of suspended particles
in a liquid)
$\vec{j}=\vec{v} \rho $ is
nonzero  in the nonequilibrium  situation and
a non-negligible  matter transport occurs as a
consequence of  the Brownian motion, on the ensemble average.

It is interesting to notice that the local velocity field
 $\vec{v}(\vec{x},t)$
  obeys the natural (local) momentum  conservation law
which   directly originates from
the rules of the It\^{o} calculus for Markovian diffusion processes,
\cite{nel}, and from the first moment equation in the
diffusion approximation  (!) of the Kramers theory, \cite{blanch}:
$${\partial _t\vec{v} + (\vec{v} \cdot \vec{\nabla }) \vec{v} =
\vec{\nabla }(\Omega - Q)\enspace .}\eqno (3)$$

An effective  potential function $\Omega (\vec{x})$
can be expressed in terms of the  forward drift
$\vec{b}(\vec{x}) = {\vec{F}(\vec{x})
\over {m\beta }}$ as follows:
$\Omega = {{\vec{F}^2} \over {2m^2\beta ^2}} + {D\over {m\beta }}
\vec{\nabla } \cdot \vec{F}$.

Let us emphasize that it is the diffusion (Smoluchowski)
approximation which makes
the right-hand-side of Eq. (3) substantially  different from the
usual moment equations appropriate for  the Brownian motion.
In particular, the force $\vec{F}$ presumed to act
upon an individual particle, does not give rise in Eq. (3)
to the  expression
$-{1\over m}\vec{\nabla }V$  which might be expected on the basis of
kinetic theory intuitions and moment identities directly derivable
from the Karmers equation, but to  the term
$+\vec{\nabla }\Omega $.

Moreover, instead of the standard pressure term,
 there appears a contribution from a
  probability density  $\rho $-dependent potential
 $Q(\vec{x},t)$. It  is given in terms of the so-called osmotic
velocity field  $\vec{u}(\vec{x},t) =
D\vec{\nabla } \, ln \rho (\vec{x},t)$, (cf. \cite{nel}):
$Q(\vec{x},t) = {1\over 2} \vec{u}^2 + D\vec{\nabla }
\cdot \vec{u}$
and  is  generic to a local momentum conservation
law  respected by   isothermal Markovian diffusion processes, cf.
 \cite{nel,vigier,blanch,zambrini}.

To  analyze   perturbations (and flows) of the noise carrier,
 a  more general function
$\vec{b}(\vec{X}(t),t)$,  must  replace the Smoluchowski drift of
Eqs. (1), (2).
The  forward drifts modify additively the pure noise
(Wiener process entry) term in the It\^{o} equations.
Under suitable restrictions, we can relate probability  measures
corresponding to  different (in terms of forward drifts !)
Fokker-Planck equations  and  processes by means of the
Cameron-Martin-Girsanov theory  of measure transformations.
The Radon-Nikodym derivative of measures is here involved and
for suitable  forward drifts which are
are gradient fields   that  yields, \cite{blanch},
the most  general  form of an auxiliary potential
$\Omega (\vec{x},t)$ in Eq. (3):
$${\Omega (\vec{x},t) = 2D[ \partial _t\phi + {1\over 2}
({\vec{b}^2\over {2D}} + \vec{\nabla }\cdot \vec{b})]\enspace .}
\eqno (4)$$
We denote $\vec{b}(\vec{x},t) = 2D \vec{\nabla } \phi (\vec{x},t)$.

Eq.  (4) is a   trivial identity,
if we take for granted that
all drifts are known from the beginning, like in case of typical
Smoluchowski diffusions where the external force $\vec{F}$ is
a priori postulated.
We can proceed otherwise and,
 on the contrary, one  can  depart
from a  suitably chosen    space-time dependent function
$\Omega (\vec{x},t)$.
From this point of view, while developing the formalism, one should
decide what is a quantity of a \it primary \rm physical interest:
the field  of drifts $\vec{b}(\vec{x},t)$ or the potential
$\Omega (\vec{x},t)$.

   Mathematical features of the formalism
appear to depend crucially on the properties (like continuity,
local and global boundedness, Rellich class) of the potential
$\Omega $, see e.g. \cite{blanch,carlen}.
Let us consider
a bounded from below (local boundedness from above is useful as well),
 continuous function $\Omega (\vec{x},t)$,
cf. \cite{blanch}).  Then,  by means of the  gradient
field ansatz for the diffusion current velocity
($\vec{v}=\vec{\nabla }S\rightarrow
\partial _t\rho = - \vec{\nabla }\cdot [(\vec{\nabla } S)\rho ]$)
we can  transform the momentum conservation law (3) of a
Markovian diffusion process to  the universal
Hamilton-Jacobi form:
$${\Omega = \partial _tS + {1\over 2} |\vec{\nabla }S|^2  + Q }
\eqno (5)$$
where $Q(\vec{x},t)$ was defined before. By
applying the  gradient operation to Eq. (5) we recover (3).
In the above,  the contribution due to $Q$
is a direct consequence of  an initial probability measure choice
for the diffusion    process,
 while $\Omega $ via Eq. (4)  does account for an appropriate
forward drift of the process.

Thus, in the context of Markovian diffusion processes,
 we can consider a  closed system of partial differential equations
 which comprises   the continuity
equation $\partial _t \rho =- \vec{\nabla }(\vec{v}\rho )$
 and  the Hamilton-Jacobi  equation
(5), plus suitable initial (and/or boundary) data.
The underlying diffusion process is specified uniquely, cf.
\cite{blanch,zambrini}.

Since the pertinent nonlinearly coupled system  equations looks
discouraging,
it is useful to mention  that a  \it
linearisation \rm of this
 problem is provided by a time-adjoint pair of
generalised diffusion equations  in the framework
 of the so-called Schr\"{o}dinger boundary data problem,
 \cite{blanch,zambrini}.
  The standard heat equation appears as a very special
  case in this formalism.

 The local conservation law  (3)  acquires a direct
 physical meaning (the
rate of change of momentum
carried by a locally co-moving with the flow volume, \cite{blanch}),
only if averaged with respect to $\rho (\vec{x},t)$ over
a   simply connected spatial area.   If $V$ stands for
a volume enclosed by a two-dimensional outward oriented
surface $\partial V$, we define a co-moving
volume on small time scales, by deforming the boundary surface in
accordance with the local current velocity  field values.
Then , let us consider  at time $t$ the  displacement of the
boundary surface $\partial V(t)$ defined  as follows:
$\vec{x}\in \partial V \rightarrow \vec{x} +
\vec{v}(\vec{x},t)\triangle t$ for all $\vec{x}\in \partial V$.
Up to the first order in $\triangle t$  this guarantees the
conservation of mass (probability measure)  contained
in $V$ at time $t$ i. e. $\int_{V(t+\triangle t)}\rho
(\vec{x},t+\triangle t)d^3x  - \int_{V(t)}\rho (\vec{x},t)d^3x
\sim 0$.

The  corresponding (to the leading order in $\triangle t$)
quantitative  momentum   rate-of-change measure reads,
cf. \cite{blanch},
$\int_V \rho \vec{\nabla }(\Omega - Q)d^3x$.

 For a particular  case of the
 free  Brownian expansion  of an initially given
 $\rho _0(\vec{x})=
{1\over {(\pi \alpha ^2)^{3/2}}} exp
(-{x^2\over \alpha ^2}) $,  where $\alpha ^2=4Dt_0$,
 we would have
 $ -\int_V\rho \vec{\nabla }Q d^3x=-
\int_{\partial V}Pd\vec{\sigma }$, where
$Q(\vec{x},t) =
{\vec{x}^2\over {8(t+t_0)^2}} - {{3D}\over {2(t+t_0)}}$, while
the "osmotic pressure" contribution reads  $P(\vec{x},t)= -{{D}
\over {2(t+t_0)}} \rho (\vec{x},t)$ for all
$\vec{x}\in R^3$ and $t \geq 0$.

The current velocity   $\vec{v}(\vec{x},t)=\vec{\nabla }
S(\vec{x},t)=
{\vec{x}\over {2(t+t_0)}}$ is  linked to the Hamilton-Jacobi
equation
$\partial _tS +{1\over 2}|\vec{\nabla }S|^2 + Q =0$
whose solution reads: $S(\vec{x},t)= {\vec{x}^2\over {4(t+t_0)}}
 + {3\over 2}D ln[4\pi D(t+t_0)]$.

Let us observe that the initial data $\vec{v}_0=-D\vec{\nabla }
ln\, \rho _0= -\vec{u}_0$ for the current
velocity field  indicate that we have totally  ignored
a crucial  \it preliminary \rm stage of the dynamics on the
$\beta ^{-1}$ time scale,  when the Brownian expansion of an
initially \it static \rm  ensemble has been ignited and so particles
have  been ultimately  set  in motion.

Notice also that our "osmotic expansion pressure" $P(\vec{x},t)$
is not positive definite, in contrast to
 the familiar kinetic theory (equation of state) expression
 for the pressure  $P(\vec{x})= \alpha \,
 \rho ^{\beta }(\vec{x}), \,  \alpha >0$ appropriate for gases.
 The admissibility of the negative sign of the "pressure"  function
 encodes the fact that the
Brownian evolving concentration of particles generically
decompresses (blows up),
instead of being compressed by the surrounding medium.

The loss  (in view of the "osmotic" migration) of momentum
 stored in a  control volume at a given time,
may be here interpreted
in terms of  an acceleration   $-\int_V\rho \vec{\nabla }Qd^3x$
induced by a \it fictituous \rm "attractive force".

By invoking an  explicit Hamilton-Jacobi connection (5),
we may attribute
to a diffusing Brownian  ensemble
 the mean kinetic energy   per unit of mass
$\int_V\rho {1\over 2}\vec{v}^2 d^3x$.
In view of $<\vec{x}^2>=6D(t+t_0)$,  we have also
$\int_{R^3}\rho {1\over 2}\vec{v}^2 d^3x= {{3D}\over {4(t+t_0)}}$.
Notice that the mean energy
$\int_V\rho ({1\over 2}\vec{v}^2+Q) d^3x$  needs not to be  positive.
Indeed, this expression identically vanishes after extending
integrations from $V$ to $R^3$.
On the other hand the kinetic contribution, initially equal
$\int_{R^3}{1\over 2} \rho v^2 d^3x = 3D/\alpha ^2$ and
evidently \it coming from nowhere, \rm continually
diminishes and is bound to disappear in the asymptotic
$t\rightarrow \infty $ limit, when  Brownian particles become
uniformly distributed in space.

Normally, diffusion processes yielding a nontrivial matter
transport (diffusion currents) are observed
for a  non-uniform  concentration of colloidal particles which
are regarded as independent (non-interacting).
We can however devise a thought (numerical) experiment
that gives rise
to a corresponding transport in terms of an ensemble of
sample (and thus independent) Brownian motion realisations on a
fixed finite  time interval,
 instead of considering a multitude of them (migrating swarm of
 Brownian particles)  simultaneously.

 Let us assume that
"an effort" (hence, an energy loss) of the random medium, on the
$\beta ^{-1}$ scale, to produce a local Brownian diffusion current
$(\rho \vec{v})(\vec{x},t_0)$  out of the  initially static ensemble
and thus to decompress (lower the blow-up tendency) an initial
non-uniform probability distribution, results in the \it effective
osmotic   reaction \rm of  the random medium.
   This  is the Brownian recoil effect of Ref. \cite{vigier}.

In that case, the particle swarm   propagation scenario becomes
entirely different from the standard  one .
First of all, the nonvanishing forward drift   $\vec{b}=\vec{u}$
is  generated as a dynamical (effective, statistical here !)
response of the bath to the enforced by the bath  particle
transport with the local
velocity $\vec{v}= -\vec{u}$.
Second, we need to account for a parellel  inversion of
the pressure effects (compression $+\vec{\nabla }Q$ should
replace the decompression $-\vec{\nabla }Q$) in the respective local
momentum conservation law.

Those features can be secured    through an explicit  realization
 of  the action-reaction principle
 which we promote to the status of the \it  third Newton law
 in the mean. \rm

On the level of Eq. (3), once averaged over a finite
volume, we interpret the momentum per unit of mass rate-of-change
$\int_V \rho \vec{\nabla }(\Omega - Q)d^3x$   which occurs
exclusively due to the Brownian expansion, to generate a
counterbalancing rate-of-change tendency in the random medium.
To account for the emerging forward drift and an obvious
modification  of the subsequent dynamics of an ensemble of
 particles, we  re-define Eq. (3) by setting
  $- \int_V \rho \vec{\nabla }(\Omega - Q)d^3x$ in its
  right-hand-side instead of  $+ \int_V \rho \vec{\nabla }
  (\Omega - Q)d^3x$ .   That amounts to   an
  instantaneous  implementation  of
  the third Newton law in the mean (action-reaction principle)
  in Eq. (3).

Hence, the momentum conservation law for the process \it
with a recoil \rm
(where the reaction term replaces the decompressive
"action" term) would read:
$${\partial _t\vec{v} + (\vec{v}\cdot \vec{\nabla })\vec{v} =
\vec{\nabla } (Q- \Omega )}\eqno (6) $$
so  that
$${\partial _t S + {1\over 2}
|\vec{\nabla }S|^2 - Q= -\Omega }\eqno (7)$$
stands for the corresponding Hamilton-Jacobi equation, cf.
\cite{zambrini,misawa}, instead of Eq. (5).
A suitable adjustment (re-setting) of the initial data is
here necessary.

In the coarse-grained picture of motion we shall deal with a
sequence    of repeatable scenarios realised
on the  Smoluchowski process  time scale $\triangle t$:
the Brownian swarm expansion build-up is accompanied by the parallel
counterflow build-up, which in turn modifies the subsequent
stage of the Brownian swarm migration  (being interpreted to modify
the forward drift of the process) and the corresponding
built-up anew counterflow.

The new closed system of partial differential equations
refers to Markovian diffusion-type  processes
again, \cite{nel,blanch,carlen}.
The link is particularly obvious if we observe that the
 new Hamilton-Jacobi equation (7) can be formally
rewritten in the previous form (5) by introducing:
$${\Omega _r= \partial _tS + {1\over 2} |\vec{\nabla }S|^2 + Q }
\eqno (8)$$
where $\Omega _r= 2Q - \Omega $ and
 $\Omega $ represents the previously defined potential function
of any Smoluchowski (or more general) diffusion process.

It  is $\Omega _r$  which via Eq. (4) would determine forward drifts
of the Markovian diffusion process with a recoil. They must obey the
Cameron-Martin-Girsanov identity
$\Omega _r = 2Q- \Omega =
2D[ \partial _t\phi + {1\over 2}
({\vec{b}^2\over {2D}} + \vec{\nabla }\cdot \vec{b})]$.

Our new closed system of equations is badly nonlinear and coupled,
but its linearisation
can be immediately given in terms of an adjoint pair of
Schr\"{o}dinger equations with  a potential $\Omega $,
\cite{nel,zambrini}.
Indeed,
$i\partial _t \psi = - D\triangle \psi + {\Omega \over {2D}}\psi $
with a solution $\psi = \rho ^{1/2} exp(iS)$
and its complex adjoint makes the job, if
we regard  $\rho $ together with $S$ to remain in conformity with
the previous notations. The choice of
$\psi (\vec{x},0)$ gives rise to a solvable Cauchy problem.
This feature we shall exploit in below.
Notice that, for time-indepedent $\Omega $, the
total energy $\int_{R^3}({v^2\over 2} -Q + \Omega )\rho d^3x$
 of the diffusing ensemble is a  conserved quantity.

  The general existence criterions for Markovian
diffusion   processes of that kind, were formulated in Ref.
\cite{carlen}, see also \cite{zambrini,blanch}.

Let us  consider a  simple one-dimensional example.
In the absence of external forces, we solve the
equations   (in  space dimension one)  $\partial _t\rho =
-\nabla (v\rho )$ and $\partial _t v + (v\nabla )v = + \nabla Q$,
with an initial probability density $\rho _0(x)$  chosen in
correspondence with the previous free Brownian motion example.
We denote $\alpha ^2= 4Dt_0$.
Then,
$\rho (x,t)=
{\alpha \over {[\pi (\alpha ^4 + 4D^2t^2)]^{1/2}}}\,
exp[-{{x^2\alpha ^2}\over  {\alpha ^4 + D^2t^2}}]$
and
$b(x,t)= v(x,t) + u(x,t)= {{2D(\alpha ^2 - 2Dt)x} \over
{\alpha ^4 +  4D^2t^2}}$
are the pertinent solutions.
Notice that $u(x,0)= -{{2Dx}\over \alpha ^2}=b(x,0)$  amounts to
$v(x,0)=0$,  while in the previous free Brownian case the initial
current velocity was equal to $-D\nabla ln\, \rho _0$.
This re-adjustment of the initial data can be interpreted in
terms of the counterbalancing (recoil) phenomenon:
the  would-be initial Brownian
ensemble current velocity  $v_0=-u_0$ is here
completely saturated by the emerging forward  drift
$b_0=u_0$, see e.g. also \cite{vigier}.
We deal also  with a fictituous "repulsive" force,
which corresponds to the compression (pressure upon) of
the Brownian ensemble due to the counter-reaction of the
surrounding  medium.
We can write things more explicitly. Namely, now:
$Q(x,t)= {{2D^2\alpha ^2}\over {\alpha ^4 + 4D^2t^2}}
({{\alpha ^2 x^2}\over {\alpha ^4+ 4D^2t^2}} - 1)$
and the corresponding pressure term ($\nabla Q=
{1\over \rho }\nabla P$) reads
$P(x,t) = - {{2D^2\alpha ^2}\over {\alpha ^4 + 4D^2t^2}}
\rho (x,t)$
giving a positive contribution $+\nabla Q$ to the local
conservation law.
The    related Hamilton-Jacobi equation
$\partial _tS + {1\over 2} |\nabla S|^2 = + Q$
is solved by
$S(x,t) = {{2D^2x^2t}\over {\alpha ^4 + 4D^2t^2}} - D\,
arctan\, (-{{2Dt}\over \alpha ^2})$.
With  the above form of $Q(x,t)$ one can readily check that
 the Cameron-Martin-Girsanov constraint euqation for the
forward drift of the Markovian diffusion process with a recoil
is automatically valid for $\phi  ={1\over 2}ln\, \rho  + S$:
$2Q = 2D[\partial _t\phi +{1\over 2}({b^2 \over {2D}} +
\nabla \cdot b)]$.

In anology with our free Brownian motion discussion, let us observe
that  presently
$<x^2> = {\alpha ^2\over 2} + {{2D^2t^2}\over \alpha ^2}$.
It is easy to demonstrate that the quadratic dependence on time
persists for arbitrarily  shaped  initial choices of the
probability distribution $\rho _0(x)>0$.
That signalizes an  anomalous behaviour (enhanced diffusion)
of the pertinent Markovian  process when $\Omega =0$
i. e. $\Omega _r=2Q$.

We can evaluate the kinetic energy contribution
$\int_{R} \rho {v^2\over 2}dx = {{4D^4t^2}
\over {\alpha ^2(\alpha ^4 + 4D^2t^2)}}$
which in contrast to the Brownian case shows up a continual growth
up to the terminal (asymptotic) value ${D^2\over \alpha ^2}$.
This value was in turn an initial kinetic  contribution
in  the previous free Brownian expansion example.
In contrast to that case, the total energy integral is now finite
(finite energy diffusions of Ref. \cite{carlen}) and reads
$\int_R({1\over 2}v^2 - Q)\rho dx = {D^2\over \alpha ^2}$
 (it  is a conservation law).
The asymptotic value of the current velocity
$v\sim {x\over t}$  is twice larger than   this appropriate for the
Brownian motion, $v\sim {x\over {2t}}$.

 It is easy to
 produce an  explicit solution  to  (7), (8)
 in case of $\Omega (x)=
{1\over 2}\gamma ^2x^2   - D\gamma $,  cf. \cite{misawa},
with  exactly the same inital probability density
$\rho _0(x)$ as before.
The forward drift of the corresponding diffusion-type process
does not show up  any obvious  contribution from the
harmonic Smoluchowski force. It
  is  completely eliminated by the Brownian recoil scenario.
One may check that  $b(x,0)= -{{2Dx}\over \alpha ^2} = u(x,0)$,
while  obviously $b={F\over {m\beta }}= -\gamma x$ would hold
true for all times, in case of the Smoluchowski
diffusion process.
Because  of the harmonic attraction  and suitable initial
probability measure choice, we have here
wiped out all  previously  discussed enhanced diffusion
features.
Now,  the dispersion  is attentuated and actually  the
non-dispersive diffusion-type  process is  realised:
$<x^2>$ does not spread at all despite of the intrinsically
stochastic  nature of the dynamics (finite-energy diffusions
of Ref. \cite{carlen}).

It is clear that stationary processes \it are the same \rm
both in case of  the  standard Brownian motion  and the
Brownian motion with  a recoil. The respective
propagation scenarios substantially  differ in the
non-stationary case only.

{\bf Acknowledgements:}
I would like to thank Professor Eric Carlen for discussion
about the scaling limits of the Boltzmann equation and  related
conservation laws.

\end{document}